\documentstyle[12pt]{article}\textwidth 16cm
\begin{document}
%
%%%%%%%%%%%%%%%%%%%%%%%%%%%%%%%%%%%%%%%%%%%%%%%%%%%%%%%%%%%%%%
%
\begin{titlepage}
\begin{raggedleft}
THES-TP 96/07\\
July 1996\\
\end{raggedleft}
\vspace{2em}
\begin{center}
{\Large\bf{ The complete Faddeev-Jackiw treatment \\ 
of the $U_{EM}(1)$ gauged SU(2) WZW model}}
\footnote {Work supported by the European Community Human
Mobility program "Eurodafne",Contract CHRX-CT92-00026.}\\
\vspace{2em}
{\large J.E.Paschalis and P.I.Porfyriadis}
\vspace{1em}\\
Department of Theoretical Physics, University of
Thessaloniki,\\GR 54006 \hspace{0.25em},\hspace{0.25em} Thessaloniki
\hspace{0.25em},\hspace{0.25em} Greece\\
\vspace{1em}
{\scriptsize PASCHALIS@OLYMP.CCF.AUTH.GR\\
PORFYRIADIS@OLYMP.CCF.AUTH.GR}
\end{center}
\vspace{2em}
\begin{abstract}
The two flavour, four dimensional WZW model coupled to electromagnetism, 
is treated as a constraint system in the context of the Faddeev-Jackiw 
approach. No approximation is made. Detailed exposition of the 
calculations is given. Solution of the constraints followed by proper 
Darboux's transformations leads to an unconstrained Coulomb-gauge 
Lagrangian density.
\end{abstract}
\end{titlepage}
%
%%%%%%%%%%%%%%%%%%%%%%%%%%%%%%%%%%%%%%%%%%%%%%%%%%%%%%%%%%%%%%%%%
%
\section{Introduction}
\label{intro}
The basic feature of the Faddeev-Jackiw (FJ) approach 
\cite{F-J,Jackiw} to constrained systems is the fact that it treats 
all the constraints on an equal basics. There is no distinction 
between first and second  class constraints as in the case of 
the Dirac quantization procedure \cite{Dirac,HRT}. 
Also constraints involving
canonical momenta $p_i$ conjugate to certain variables $q_i$ ,
 whose velocities $\dot{q_i}$ occur lineary or do not occur at
 all in the Lagrangian, are absent in the FJ approach 
 contrary to the Dirac procedure.

The FJ approach is based on the Darboux's theorem
 \cite{Jackiw,Arnold}, according to 
which an arbitrary vector potential $a_{i}(\xi)$, whose
associated field strength $f_{ij}(\xi)$ is non-singular can be 
written apart from a gauge transformation in the form  
\begin{equation}
  a_i(\xi) = \frac{1}{2}Q^k(\xi)\omega_{kl}
      \frac{\partial Q^{l}(\xi)}{\partial \xi^i}\; \; .
\end{equation}
and the field strength as
\begin{equation}
  f_{ij}(\xi) = \frac{\partial Q^{k}(\xi)}{\partial \xi^i}
  \omega_{kl}\frac{\partial Q^{l}(\xi)}{\partial \xi^j}\; \; .
\end{equation}
where $ Q^{i}(\xi) $ are the new (Darboux transformed) coordinates
and $ [\omega_{ij}] $ is a constant, non-singular antisymmetric matrix.
In this case the associated canonical one form 
$ a_i(\xi)d\xi^i$ is called diagonal.

The starting point of the FJ approach is a Lagrangian first order 
in the time derivatives. Any conventional, second order in time 
derivatives Lagrangian can be written as a first order expression by
properly enlarging its configuration space so that it includes 
the conjugate momenta of the coordinate variables. For a constrained 
system described by the Lagrangian
\begin{equation}
   {\cal L}(\xi,\dot{\xi}) = a_i(\xi)\dot{\xi^i} -
                             H(\xi)\;\; , \;\;  i=1,...,N
\end{equation}
the field strength  $f_{ij}=\frac{\partial}{\partial \xi_i}
a_{j}(\xi) -\frac{\partial}{\partial \xi_j}a_{i}(\xi) $
is a singular $N\times N$ matrix. The Darboux's transformations 
can be applied to the maximal non-singular submatrix of 
$f_{ij}$. The Lagrangian tranforms into
\begin{equation}
  {\cal L}(Q,\dot{Q},z)=\frac{1}{2}Q^{k}(\xi)
  \omega_{kl}\dot{Q^l}(\xi) -
                        H(Q,z)\;\; , \;\; k,l=1,...,2n
\end{equation}
where by $z$ we denote the coordinates left unchanged. Then by
using the Euler-Lagrange equations we solve for as many $z$'s as 
possible in terms of other coordinates and we obtain 
\begin{equation}
  {\cal L}(Q,\dot{Q},z)=\frac{1}{2}Q^{k}(\xi)\omega_{kl}
  \dot{Q^l}(\xi)-H'(Q)-z^m\Phi_m(Q)\; \; ,
\end{equation}
where $\Phi_m$ are the true constraints of the system. Next we solve
the constraint equations $\Phi_{m}(Q)=0$ for some of the Q's and
substitute back  in (5). We end up with a Lagrangian with the same
form as the initial (3) and with fewer dynamical variables. The whole
 procedure is then repeated again until we are left with an  
 unconstrained Lagrangian, with reduced coordinate space 
 and with diagonal canonical one-form. In the case that  
 the constraints cannot be solved due to technical difficulties 
 one can return to the Dirac procedure.

The equivalence of the FJ approach to the Dirac method 
is discussed in \cite{Govaerts,Montani}. Its extension in superspace 
is given in \cite{Kulsh1,Barcelos}. 
Application of the approach to the light-cone quantum field theory 
is given in \cite{Jun,Jacob}, to non-abelian systems in \cite{MW} 
to hidden symmetries in \cite{Wotzasek} and to self-dual fields in 
\cite{Kulsh2}. Other works related to the FJ 
approach are given in \cite{BW2,Cronstrom,PP2}. 
%
%%%%%%%%%%%%%%%%%%%%%%%%%%%%%%%%%%%%%%%%%%%%%%%%%%%%%%%%%%%%%%%%%%%
%
\section{The $U_{EM}(1)$ gauged SU(2) WZW model in the FJ framework}
\label{WZW}

The model \cite{Witten,Pak,Balachandran,Donoghue} 
describes the low energy interactions of pions and photons
including those related to the QCD anomalies. It posesses the 
symmetries of electromagnetism and those relevant to QCD and has no 
extra symmetries. The effective action is given by 
\begin{eqnarray}
  \Gamma_{eff}(U,A_\mu)&\!\!\!\! =&\!\!\!\! {\Gamma}_{EM} (A_\mu) +
                       \Gamma_\sigma (U,A_\mu) +
                       {\Gamma}_{WZW} (U,A_\mu)\; \; ,\\
  {\Gamma}_{EM} (A_\mu)&\!\!\!\!=&\!\!\!\! -\frac{1}{4}\int\!d^4\!x
                 F_{\mu \nu} F^{\mu \nu}\; \; ,\nonumber \\
  \Gamma_\sigma (U,A_\mu)&\!\!\!\! =&\!\!\!\! -\frac{f_\pi^2}{16}
                 \int\!d^4\!x \mbox{tr}\,(R_\mu R^\mu)\nonumber \\
                 &\!\!\!\!=&\!\!\!\!-\frac{f_\pi^{2}}{16}
                 \int\!d^4\!x \mbox{tr}\,(r_\mu r^\mu) +
                 \frac{if_\pi^{2}e}{8}
                 \int\!d^4\!x A_\mu \mbox{tr}\,[Q(r_\mu-l_\mu)]
                 \nonumber \\ &\!\!\!\!+&\!\!\!\! \frac{f_\pi^{2}e^2}{8}
                 \int\!d^4\!x A_\mu A^\mu \mbox{tr}\,
                 (Q^2-U^\dagger QUQ)\; \; , \nonumber \\
  {\Gamma}_{WZW} (U,A_\mu)&\!\!\!\! =&\!\!\!\! -\frac{N_{c}e}{48\pi^2}
                 \int\!d^4\!x \epsilon^{\mu \nu \alpha \beta} 
                  A_\mu \mbox{tr}\,[Q(r_\nu r_\alpha r_\beta +
                  l_\nu l_\alpha l_\beta)]\nonumber \\
                  &\!\!\!\!+&\!\!\!\! \frac{iN_{c}e^2}{24\pi^2}
                  \int\!d^4\!x \epsilon^{\mu \nu \alpha \beta} 
                  A_\mu(\partial_\nu A_\alpha)
                  \mbox{tr}\,[Q^2(r_\beta+l_\beta) \nonumber \\
                  &\!\!\!\!+&\!\!\!\! \frac{1}{2}QU^\dagger QUr_\beta
                 +\frac{1}{2}QUQU^\dagger l_\beta] \nonumber \; \; ,
\end{eqnarray}
where
\begin{eqnarray}
   U&\!\!\!\!=&\!\!\!\!\exp\,(2i\theta_i\tau_i/f_\pi)\;\; ,\hspace{1em}
   r_\mu=U^\dagger\partial_\mu U\; \; ,\hspace{1em}
   R_\mu=U^\dagger D_\mu U\; \; ,\nonumber \\
   l_\mu&\!\!\!\!=&\!\!\!\!(\partial_\mu U)U^\dagger \; \; ,\hspace{1em}
   L_\mu=(D_\mu U)U^\dagger\; \; .\nonumber 
\end{eqnarray}
See Appendix for notation.

In the SU(3) case the effective action should include one more term 
\begin{eqnarray}
   {\Gamma}_{WZW}(U)=-\frac{iN_c}{240\pi^2}\int\!d^5\!x\epsilon^{ijklm}
                     \mbox{tr}\,(l_{i}l_{j}l_{k}l_{l}l_{m})\; \; ,
\end{eqnarray}
which describes processes like $K^{+}K^{-}\rightarrow \pi^{+}\pi^{0}
\pi^{-}$ (in the low energy regime) which respect the combined discrete 
operations $ U(\bf x \rm,t)\rightarrow U(-\bf x\rm,t)\;,\; 
U(\bf x \rm,t)\rightarrow U^{\dagger}(\bf x\rm,t)$ but not each one 
seperately. In the SU(2) case this term vanishes identically. 
Nevertheless the term ${\Gamma}_{WZW} (U,A_\mu)$ which comes from 
gauging (7) is present and describes processes like 
$\pi^{0}\rightarrow 2\gamma\;,\; \gamma \rightarrow 3\pi\;,\;
2\gamma \rightarrow 3\pi\;$ etc. related to the QCD anomaly.

In \cite{PP} the U field was expanded in powers of the Goldstone boson 
fields $\theta_a $. Keeping terms up to second and third order 
we obtained Lagrangian densities with only one true constraint, 
the one which is multiplied by $ A_0$ (scalar potential). Then 
after solving the equation of the constraint and performing 
the proper Darboux's
transformations the longitudinal part of the vector potential was 
cancelled out and we were left with an unconstrained Coulomb-gauge 
Lagrangian density. This is exactly what happens also in the case of 
spinor electrodynamics \cite{F-J}. 

In this work no expansion is made. Instead a complete treatment 
is given to the model as a constrained system in the context of 
the FJ formalism. The  
following parametrization of the U field is used 
\begin{eqnarray}
   U&\!\!\!\!=&\!\!\!\!
   \cos(2|\theta|/f_\pi)+i\tau_i \hat \theta_i \sin(2|\theta|/
   f_\pi)=\phi_{0}(2|\theta|/f_\pi)
   +i\tau_i \phi_{i}(2|\theta|/f_\pi)\; \; , 
\end{eqnarray}
where the real fields $ \phi_0 \;,\;\phi_i \;\; ,\;\; i=1,2,3 $ 
are subject to the unitarity constraint
\begin{eqnarray}
\phi_{0}^2+\phi_{i}^2=1
\end{eqnarray}
The Lagrangian density of the model is written as follows in terms 
of the $\phi$-fields
\begin{eqnarray}
  {\cal L}_{eff}&\!\!\!\!=&\!\!\!\!{\cal L}_{EM}+
                 {\cal L}_{\sigma}+{\cal L}_{WZW} ,\\
  {\cal L}_{EM}&\!\!\!\!=&\!\!\!\!-\frac{1}{4}
                F_{\mu \nu}F^{\mu \nu}\; \; ,\nonumber \\
  {\cal L}_{\sigma}&\!\!\!\!=&\!\!\!\! \frac{f_{\pi}^2}{8}
               \partial_\mu \phi_{0}\partial^\mu\phi_{0}+
	       \frac{f_{\pi}^2}{8}
               \partial_\mu \phi_{i}\partial^\mu\phi_{i}+
	       \frac{f_{\pi}^2 e}{4}
               A^\mu(\phi_{2}\partial_\mu\phi_{1}-
                      \phi_{1}\partial_\mu\phi_{2})
               +\frac{f_{\pi}^2 e^2}{8}A_\mu A^\mu
                (\phi_{1}^{2}+\phi_{2}^{2})\; \; ,\nonumber \\
  {\cal L}_{WZW}&\!\!\!\!=&\!\!\!\!
               -\frac{2N e}{\phi_0}
               {\epsilon^{\mu \nu \alpha \beta}}A_\mu\partial_\nu
               \phi_1\partial_\alpha \phi_2\partial_\beta
	       \phi_3 \; \; \nonumber \\
	        &\!\!\!\!+&\!\!\!\!2N e^2
		\phi_3{\epsilon^{\mu \nu \alpha \beta}}A_\mu
		(\partial_\nu A_\alpha)\partial_\beta \phi_0
	       - N e^2\phi_0
		\phi_3{\epsilon^{\mu \nu \alpha \beta}}
		(\partial_\mu A_\nu)
		(\partial_\alpha A_\beta) \;\; , \nonumber
\end{eqnarray}
where $ N=\frac{N_c}{24\pi^2}\;.\; $
The canonical momenta conjugate to $\phi_0 \;,\;\phi_i \;$ 
are given by
\begin{eqnarray}
 \vspace{0.5em}
  p_{0}&\!\!\!\!=&\!\!\!\!\frac{\partial {\cal L}_{eff}}
                 {\partial \dot{\phi}_0}=
         \frac{f_{\pi}^2}{4}\dot{\phi}_{0}-
         2N e^2(\bf A\cdot B\rm)\phi_3\;\;,\nonumber \\
\vspace{0.5em}
\vspace{0.5em}
  p_{1}&\!\!\!\!=&\!\!\!\!\frac{\partial {\cal L}_{eff}}
                 {\partial \dot{\phi}_1}=
        \frac{f_{\pi}^2}{4}\dot{\phi}_{1}
	+\frac{f_{\pi}^2e}{4}A_{0}\phi_{2}
	-\frac{ 2N e}{\phi_0}(\nabla\phi_2
	\times\nabla\phi_3)\cdot \bf A\;\;,\\
\vspace{0.5em}
  p_{2}&\!\!\!\!=&\!\!\!\!\frac{\partial {\cal L}_{eff}}
                 {\partial \dot{\phi}_2}=
        \frac{f_{\pi}^2}{4}\dot{\phi}_{2}
	-\frac{f_{\pi}^2e}{4}A_{0}\phi_{1}
	+\frac{2N e}{\phi_0}(\nabla\phi_1
	\times\nabla\phi_3)\cdot \bf A\;\;,\nonumber \\
\vspace{0.5em}
  p_{3}&\!\!\!\!=&\!\!\!\!\frac{\partial {\cal L}_{eff}}
                 {\partial \dot{\phi}_3}=
        \frac{f_{\pi}^2}{4}\dot{\phi}_{3}
	-\frac{2N e}{\phi_0}(\nabla\phi_1
	\times\nabla\phi_2)\cdot \bf A\;\;.\nonumber 
\end{eqnarray}
In the enlarged configuration space with coordinates
$-\mbox{\boldmath $\pi$} \; , \; \bf A\rm 
\; , \; p_0 \; , \; \phi_0 \; , \; p_i \; , \; \phi_i \; , \; 
i=1,2,3 \; $ the
effective Lagrangian density (10) is written as an expression 
first order in time derivatives as follows 
\begin{eqnarray}
  {\cal L}_{eff}&\!\!\!\!=&\!\!\!\!-\mbox{\boldmath $\pi$}
                  \bf\cdot\dot{A}\rm+
                 p_0\dot{\phi_0}+ p_i\dot{\phi_i}-H^{0}
                -A_{0}(\rho_\sigma+\rho_{\rm w}-
		\nabla\cdot \mbox{\boldmath $\pi$})
                \; \; ,\\
\vspace{1em}
      H^{0} &\!\!\!\!=&\!\!\!\!\frac{1}{2}
               [{\mbox{\boldmath $\pi$}}^2+\bf B^{\rm 2}\rm]
                         \nonumber\\
   &\!\!\!\!+&\!\!\!\!
                 \frac{f_{\pi}^2}{8}
               (\nabla \phi_{0})^2+
	       \frac{f_{\pi}^2}{8}
               (\nabla \phi_{i})^2+
	       \frac{f_{\pi}^2 e}{4}
               \bf{A}\cdot\rm(\phi_{1}\nabla\phi_{2}-
                      \phi_{2}\nabla\phi_{1})
               +\frac{f_{\pi}^2 e^2}{8}{\bf A}^{\rm 2}\rm
                (\phi_{1}^{2}+\phi_{2}^{2}) \nonumber \\
    &\!\!\!\!+&\!\!\!\!\frac{2}{f_{\pi}^2}
              [p_i+\frac{N e}{\phi_0}
	      \epsilon_{ijk}\bf A \rm \cdot(\nabla\phi_j
	      \times\nabla\phi_k)]^2+
	      \frac{2}{f_{\pi}^2}
              [p_0+2N e^2
	      \phi_3(\bf A \cdot B \rm)]^2
                        \;\; \nonumber\\ 
   &\!\!\!\!-&\!\!\!\!2N e^2 \phi_3
        (\bf A\times \mbox{\boldmath $\pi$}\rm)\cdot \nabla \phi_0 
	-2N e^2\phi_0 \phi_3
	(\mbox{\boldmath $\pi$}\cdot \bf B\rm) \; \; ,\nonumber \\
\vspace{1em}
  \rho_\sigma&\!\!\!\!=&\!\!\!\! e(p_{2}\phi_1-p_{1}\phi_{2})
                 \; \; ,\nonumber \\
 \rho_{\rm w}&\!\!\!\!=&\!\!\!\!2N e^2 
          \nabla \phi_0 \cdot \nabla\times(\phi_3\bf A \rm)+
	 \frac{2N e}{\phi_0}
	  (\nabla\phi_1\times\nabla\phi_2)\cdot\nabla\phi_3
	   \; \; .\nonumber		  
\end{eqnarray}
We see that the canonical one form in (12) can be written apart 
from a total time derivative, in the diagonal form defined in (1).
Taking the time derivative of the unitarity constraint (9) and 
substituting the  time derivatives of the $\phi$-fields from
(11) we obtain the following secondary constraint
\begin{eqnarray}
p_0\phi_0+p_i\phi_i+\frac{N e}{\phi_0}\epsilon_{ijk}
\phi_i (\nabla\phi_j\times\nabla\phi_k)\cdot\bf A\rm+
2N e^2 \phi_0 \phi_3 (\bf A\cdot B\rm)=0
\end{eqnarray}
Setting the time derivative of (13) equal to zero does not give any
new constraint. One more constraint appears in (12) multiplied 
by $A_0$
\begin{eqnarray}
\nabla\cdot \mbox{\boldmath $\pi$}-\rho_{\sigma}-\rho_{\rm w}=0
\end{eqnarray}
whose time derivative also does not add any new one to the model. So 
finally we have three true second class constraints given by 
the equations (9),(13),(14). In order to solve (14)
we decompose the electric field $\mbox{\boldmath $\pi$}$
and the vector potential $\bf A$ into transverse and
longitudinal components
\begin{displaymath}
 \bf A^{\rm T}=\bf A\rm -\nabla A^{L'}
             \;\; , \;\; \hspace{1em}
     \bf A^{\rm L}\rm=\nabla A^{L'}
            \;\; , \;\; \hspace{1em}
      A^{\rm L'}\rm=\frac{1}{\nabla^2}
(\nabla \cdot \bf A)\rm\;\; , \;\;
\end{displaymath}
\begin{displaymath}
\mbox{\boldmath $\pi$}^{T}=\mbox{\boldmath $\pi$}
-\frac{\nabla}{\nabla^2} \pi^{L'} \;\; , \;\; \hspace{1em}
\mbox{\boldmath $\pi$}^{L}=\frac{\nabla}{\nabla^2} \pi^{L'}\;\;
, \;\; 
\hspace{1em}
      \pi^{L'}=
\nabla \cdot \mbox{\boldmath $\pi$}\rm\;\; . \;\;
\end{displaymath}

Then (14) implies that
\begin{equation}
   \mbox{\boldmath $\pi$}^L=\frac{\bf \nabla \rm}{\bf \nabla \rm^2}
   (\rho_\sigma+\rho_{\rm w})\; \; .
\end{equation}
We substitute into (12) and we obtain, apart from a total spatial 
derivative, the following expression for the effective Lagrangian 
density
\begin{eqnarray}
  {\cal L}_{eff} &\!\!\!\!=&\!\!\!\! 
  -\mbox{\boldmath $\pi$}^T\cdot\dot{\bf A}^{\rm T}\rm
       +p_0\dot{\phi}_0 +p_i\dot{\phi}_i
       +(\rho_\sigma+\rho_{\rm w})\frac{\bf \nabla \rm}
        {\bf \nabla \rm^2}\cdot\dot{\bf A}^{\rm L}\rm
	-H^{0} \;\; ,
\end{eqnarray}
where the expression for $H^0$ is given in (12) with 
$\mbox{\boldmath $\pi$}^L $ substituted by (15).
We see that $\bf A^{\rm L}$ enters the canonical one form in an 
uncanonical way. A partial diagonalization can be achieved by 
performing the following Darboux's transformations
\vspace{1em}
\begin{displaymath}
 p_1\rightarrow p_1 \cos{\alpha}+p_2\sin{\alpha} 
             \;\; , \;\; \hspace{1em}
 \phi_1 \rightarrow \phi_1 \cos{\alpha}+\phi_2 \sin{\alpha}
             \;\; , \;\;
\end{displaymath}
\begin{equation}
 p_2 \rightarrow p_2 \cos{\alpha}-p_1\sin{\alpha}
             \;\; , \;\; \hspace{1em}
 \phi_2 \rightarrow \phi_2 \cos{\alpha}-\phi_1 \sin{\alpha}
             \;\; , \;\;
\end{equation}
where 
$\alpha = e\frac{\bf \nabla \rm}{\bf \nabla \rm^2}\cdot\bf A^{\rm L}$

Then the effective Lagrangian density acquires the form 
\begin{eqnarray}
  {\cal L}_{eff} &\!\!\!\!=&\!\!\!\! 
  -\mbox{\boldmath $\pi$}^T\cdot\dot{\bf A}^{\rm T}\rm
                         +p_0\dot{\phi}_0+p_i\dot{\phi}_i
       +\rho_{\rm w} \frac{\bf \nabla \rm}
        {\bf \nabla \rm^2}\cdot\dot{\bf A}^{\rm L}\rm
	-H^{0}\;\; .
\end{eqnarray}
We have redefined $H^0$ after having performed the 
Darboux's transformations (17) into its previous expression in 
(16). Also $\rho_{\rm w} $ given initially in (12) becomes
\begin{eqnarray}
 \rho_{\rm w}&\!\!\!\!=&\!\!\!\!2N e^2 
          \nabla \phi_0 \cdot \nabla\times(\phi_3\bf A^{\rm T} \rm)+
	 \frac{2N e}{\phi_0}
	  (\nabla\phi_1\times\nabla\phi_2)\cdot\nabla\phi_3
	   \; \; .\nonumber		  
\end{eqnarray}
We see that $\bf A^{\rm L}$ cancels out. This is important for the 
next step which is the elimination of 
$\rho_{\rm w} \frac{\bf \nabla \rm}
        {\bf \nabla \rm^2}\cdot\dot{\bf A}^{\rm L}\rm $ from 
the canonical one-form in (18). Apart from total derivatives we have 
\begin{eqnarray}
  {\cal L}_{eff} &\!\!\!\!=&\!\!\!\! 
  -[\mbox{\boldmath $\pi$}^T
  -2N e^2 (\nabla\phi_0\times\bf A^{\rm L}\rm) \phi_3]
  \cdot\dot{\bf A}^{\rm T}\rm +
   [p_0 + 2N e^2 \bf A^{\rm L}\rm\cdot\nabla\times
   (\phi_3 \bf A^{\rm T}\rm)] \dot{\phi_0} \nonumber \\
   &\!\!\!\!+&\!\!\!\![p_i+\frac{N e}{\phi_0}
   \epsilon_{ijk}
   (\nabla\phi_j\times\nabla\phi_k)\cdot\bf A^{\rm L}\rm
   +2N e^2 \delta_{i3}
   (\nabla\phi_0\times\bf A^{\rm L}\rm)
   \cdot\bf A^{\rm T}]\dot{\phi_i}\rm-H^{0}
\end{eqnarray}	
We perform the second set of Darboux's transformations
\begin{displaymath}
   \mbox{\boldmath $\pi$}^T \rightarrow \mbox{\boldmath $\pi$}
   +2N e^2 (\nabla\phi_0\times\bf A^{\rm L}\rm)
   \phi_3\;\; , \;\;
\end{displaymath}
\begin{eqnarray}
   p_0 \rightarrow p_0 - 2N e^2\bf A^{\rm L}\rm\cdot 
   \nabla\times(\phi_3 \bf A^{\rm T}\rm)\;\; , \;\;
\end{eqnarray}
\begin{displaymath}
   p_i \rightarrow p_i-\frac{N e}{\phi_0}
   \epsilon_{ijk}
   (\nabla\phi_j\times\nabla\phi_k)\cdot\bf A^{\rm L}\rm
   -2N e^2 \delta_{i3}
   (\nabla\phi_0\times\bf A^{\rm L}\rm)
   \cdot\bf A^{\rm T}\rm \;\; , \;\; i=1,2,3
\end{displaymath}
which diagonalizes completely the canonical one-form in (19). 
We note that the Darboux tranformed field 
$\mbox{\boldmath $\pi$}$ is no longer tranverse. So the full
expression for the effective Lagrangian density (12) transforms as 
follows under the first two sets of Darboux's transformations (11) 
and (20)
\begin{eqnarray}
  {\cal L}_{eff}&\!\!\!\!=&\!\!\!\!-\mbox{\boldmath $\pi$}
                  \bf\cdot\dot{A}^{\rm T}\rm+
                 p_0\dot{\phi_0}+p_i\dot{\phi_i}-H^{0}
                 -H(\bf A^{\rm L}\rm)
                \; \; ,\\
\vspace{1em}
   H^{0} &\!\!\!\!=&\!\!\!\!\frac{1}{2}
           [{\mbox{\boldmath $\pi$}}^2
	    -\rho^{*}\frac{1}{\bf \nabla \rm^2}\rho^{*} +\bf B^2\rm]
                        \nonumber\\
   &\!\!\!\!+&\!\!\!\!
                 \frac{f_{\pi}^2}{8}
               (\nabla \phi_{0})^2+
	       \frac{f_{\pi}^2}{8}
               (\nabla \phi_{i})^2+
	       \frac{f_{\pi}^2 e}{4}
               \bf{A}^{\rm T}\cdot\rm(\phi_{1}\nabla\phi_{2}-
                      \phi_{2}\nabla\phi_{1})
               +\frac{f_{\pi}^2 e^2}{8}({\bf A}^{\rm T})^{\rm 2}\rm
                (\phi_{1}^{2}+\phi_{2}^{2})  \nonumber \\
   &\!\!\!\!+&\!\!\!\!\frac{2}{f_{\pi}^2}
             [p_i+\frac{N e}{\phi_0}
	      \epsilon_{ijk}\bf A^{\rm T} \rm \cdot(\nabla\phi_j
	      \times\nabla\phi_k)]^2+
	      \frac{2}{f_{\pi}^2}
             [p_0+2N e^2
	      \phi_3(\bf A^{\rm T} \cdot B \rm)]^2
                         \nonumber\\ 
   &\!\!\!\!-&\!\!\!\! 2N e^2 \phi_3
	(\mbox{\boldmath $\pi$} + \frac{\nabla}{\nabla^2}\rho^{*})
	\cdot \nabla\times(\phi_0 \bf A^{\rm T}\rm)
	-\nabla\cdot\mbox{\boldmath $\pi$}\frac{1}{\nabla^2}\rho^{*}  
	 \; \; ,\nonumber \\
   H(\bf A^{\rm L}\rm)&\!\!\!\!=&\!\!\!\!
                 -2N^2 e^4 \phi_{3}^2 (\nabla\phi_0\times
                 \bf A^{\rm L}\rm)^2 -4N^2 e^4 \phi_{3}^2
               (\nabla\phi_0\times\bf A^{\rm L}\rm)\cdot \nabla
                 \times(\phi_0 \bf A^{\rm T}\rm) \nonumber \\
   &\!\!\!\!-&\!\!\!\! \frac{8 N^2 e^4}{f_{\pi}^2 \phi_{0}^2}	
      [(\nabla\phi_3\times\bf A^{\rm T}\rm)\cdot\bf A^{\rm L}\rm]^2
                     \; \; ,\nonumber \\
\vspace{1em}
   \rho^{*}&\!\!\!\!=&\!\!\!\!\rho_\sigma +\rho_{\rm w} -
     \nabla\cdot\mbox{\boldmath $\pi$}\;\; , \nonumber \\
  \rho_\sigma&\!\!\!\!=&\!\!\!\! e(p_{2}\phi_1-p_{1}\phi_{2})
                 \; \; ,\nonumber \\
 \rho_{\rm w}&\!\!\!\!=&\!\!\!\!2N e^2
          \nabla \phi_0 \cdot \nabla\times(\phi_3\bf A^{\rm T} \rm)+
	  \frac{2N e}{\phi_0}
	  (\nabla\phi_1\times\nabla\phi_2)\cdot\nabla\phi_3
	              \; \; . \nonumber 
\end{eqnarray}
In order to obtain (21) we made use of all three constraints. 
It is interesting to note that the unitarity constraint (9) 
does not change under the first two sets of Darboux's transformations 
while the second one in (13) transform into 
\begin{displaymath}
p_0\phi_0+p_i\phi_i+\frac{N e}{\phi_0}\epsilon_{ijk}
\phi_i (\nabla\phi_j\times\nabla\phi_k)\cdot\bf A^{\rm T}\rm+
2N e^2 \phi_0 \phi_3 (\bf A^{\rm T}\cdot B\rm)
-\frac{2N e^2}{\phi_0}(\bf A^{\rm T}\rm  \times \bf 
A^{\rm L}\rm)\cdot\nabla\phi_3 =0
\end{displaymath}
and it is this form of the constraint that was actually used.
We see in (21) that although $\bf A^{\rm L} $ disappears from the 
canonical one-form, still appears in the Hamiltonian density and 
actually that part which comes from the Wess-Zumino term, contrary 
to what happens 
in \cite{PP} where the U-field was expanded in series of powers of 
the pion fields.

The $\phi$-fields are not independent due to the unitarity constraint
(9). So we use (9) to express $\dot{\phi_0}$ in terms of the other 
$\phi$-fields. Replasing into (21) we see that the canonical one-form 
loses its diagonal form
\begin{eqnarray}
  {\cal L}_{eff} &\!\!\!\!=&\!\!\!\! 
  -\mbox{\boldmath $\pi$}\cdot\dot{\bf A}^{\rm T}\rm
  +(p_i - \frac{p_0}{\phi_0}\phi_i)\dot{\phi_i} -H^{0}
  -H(\bf A^{\rm L}\rm)\;\; .
\end{eqnarray}
Rediagonalization can be realized by performing the following set of 
 Darboux's transformations
\begin{displaymath}
  p_i \rightarrow p_i+\frac{p_0}{\phi_0}\phi_i \;\;,\;\;\;\; i=1,2,3
\end{displaymath}
Next we decompose $\mbox{\boldmath $\pi$}$ into longitudinal and 
transverse components. We obtain after cancellations the following 
expression for the effective Lagrangian density
\begin{eqnarray}
  {\cal L}_{eff}&\!\!\!\!=&\!\!\!\!-\mbox{\boldmath $\pi$}^T
                  \bf\cdot\dot{A}^{\rm T}\rm+
                 p_i\dot{\phi_i}-H^{0}
                 -H(\bf A^{\rm L}\rm)
		 -H(p_0)
                \; \; ,\\
\vspace{1em}     
   H^{0} &\!\!\!\!=&\!\!\!\!\frac{1}{2}
               [({\mbox{\boldmath $\pi$}}^T)^2
	       -\rho\frac{1}{\bf \nabla \rm^2}\rho +\bf B^2\rm]
                        \nonumber\\
    &\!\!\!\!+&\!\!\!\!
                 \frac{f_{\pi}^2}{8}
               (\nabla \phi_{0})^2+
	       \frac{f_{\pi}^2}{8}
               (\nabla \phi_{i})^2+
	       \frac{f_{\pi}^2 e}{4}
               \bf{A}^{\rm T}\cdot\rm(\phi_{1}\nabla\phi_{2}-
                      \phi_{2}\nabla\phi_{1})
               +\frac{f_{\pi}^2 e^2}{8}({\bf A}^{\rm T})^{\rm 2}\rm
                (\phi_{1}^{2}+\phi_{2}^{2}) \nonumber \\
   &\!\!\!\!+&\!\!\!\!\frac{2}{f_{\pi}^2}
             [p_i+\frac{N e}{\phi_0}
	      \epsilon_{ijk}\bf A^{\rm T} \rm \cdot(\nabla\phi_j
	      \times\nabla\phi_k)]^2+
	      \frac{8 N^2 e^4}{f_{\pi}^2}
	      \phi_3^2 (\bf A^{\rm T} \cdot B \rm)^2
                         \nonumber\\ 
   &\!\!\!\!-&\!\!\!\!2N e^2 \phi_3
	(\mbox{\boldmath $\pi$}^T + \frac{\nabla}{\nabla^2}\rho)
	\cdot \nabla\times(\phi_0 \bf A^{\rm T}\rm)  
	 \; \; ,\nonumber \\
   H(p_0)&\!\!\!\!=&\!\!\!\!\frac{2}{f_{\pi}^2 }
   \frac{p_{0}^2}{\phi_{0}^2}
             +\frac{4}{f_{\pi}^2}\frac{p_0}{\phi_0}
             [p_i \phi_i +\frac{N e}{\phi_0}
	      \epsilon_{ijk}\phi_i(\nabla\phi_j
	\times\nabla\phi_k)\cdot \bf A^{\rm T} \rm+2 N e^2
	      \phi_0 \phi_3(\bf A^{\rm T} \cdot B \rm)]
                        \;\; , \nonumber \\ 
	\rho &\!\!\!\!=&\!\!\!\! \rho_\sigma +\rho_{\rm w} \;\; ,
	\nonumber			   		  
\end{eqnarray}
where $H(\bf A^{\rm L}\rm)\;,\;\rho_\sigma \;,\; \rho_{\rm w} $ is 
given in (21).
The fact that the variables $\bf A^{\rm L}\rm$ and $p_0$ 
which are absent in the 
canonical one-form still occur in the Hamiltonian  
leads to a new set of constraint equations.
\begin{equation}
\frac{\partial H(\bf A^{\rm L}\rm)}{\partial \bf A^{\rm L}\rm}=0
\;\;\;\; , \;\;\;\;  
\frac{\partial H(p_0)}{\partial p_0}=0
\end{equation}
the solutions of which are given by 
\begin{eqnarray}
\bf A^{\rm L}\rm &\!\!\!\!=&\!\!\!\!-\bf A^{\rm T}\rm +
 \frac{\phi_0}{(\nabla\phi_0)^2}
 \frac{(\bf A^{\rm T}\rm\times\nabla\phi_3)\times[\nabla\phi_0
 \times(\bf B\rm\times\nabla\phi_0)]}
 {(\nabla\phi_3\times\nabla\phi_0)\cdot \bf A^{\rm T}\rm}
 \;\; , \\
p_0 &\!\!\!\!=&\!\!\!\!-[p_i \phi_i + 
 \frac{N e}{\phi_0}\epsilon_{ijk}
 \phi_i(\nabla\phi_j\times\nabla\phi_k)\cdot \bf A^{\rm T} \rm
 +2N e^2 \phi_0 \phi_3(\bf A^{\rm T} 
 \cdot B \rm)]\phi_0  \;\; . \nonumber
\end{eqnarray}
Note also that
\begin{displaymath}
   det \left[\frac{\partial H(\bf A^{\rm L}\rm)}
   {\partial A^{L}_i \partial A^{L}_j}\right]=
   -\frac{256 N^6 e^{12}}{f^{2}_{\pi}}\frac{\phi_{3}^4}{\phi_{0}^2}
   (\nabla \phi_0)^2 [\nabla\phi_0\cdot(\nabla\phi_3\times
   \bf A^{\rm T}\rm)]^2 \not=0
\end{displaymath}
So finally after substituting (25) into (23) we are left with 
an unconstrained Lagrangian density 
with reduced coordinate space and diagonal canonical one-form 
given by
\begin{eqnarray}
  {\cal L}_{eff}&\!\!\!\!=&\!\!\!\!
  -\mbox{\boldmath $\pi$}^T\cdot\dot{\bf A}^{\rm T}+
             \rm p_{i}\dot{\phi_{i}} - H_{C}\; \; ,\\
   H_{C} &\!\!\!\!=&\!\!\!\!\frac{2}{f_{\pi}^2}
           [p_i+\frac{N e}{\phi_0}
	      \epsilon_{ijk}\bf A^{\rm T} \rm \cdot(\nabla\phi_j
	      \times\nabla\phi_k)]^2+
	      \frac{8 N^2 e^4}{f_{\pi}^2}
	      \phi_{3}^2(\bf A^{\rm T} \cdot B \rm)^2
	      \nonumber \\
          &\!\!\!\!-&\!\!\!\! \frac{2}{f_{\pi}^2}
        [p_i \phi_i+\frac{N e}{\phi_0}
	      \epsilon_{ijk}\phi_i(\nabla\phi_j
	      \times\nabla\phi_k)\cdot\bf A^{\rm T} \rm + 
              2 N e^2 \phi_0
	      \phi_3(\bf A^{\rm T} \cdot B \rm)]^2
                     \nonumber \\
          &\!\!\!\!+&\!\!\!\!\
	   \frac{1}{2}[(\mbox{\boldmath $\pi$}^T)^2
                  +\bf B^{\rm 2}\rm-
                  \rho\frac{1}{\bf \nabla \rm^2}
                  \rho+\frac{f_{\pi}^2}{4}(\nabla\phi_{0})^{2}+
		  \frac{f_{\pi}^2}{4}(\nabla\phi_{i})^{2}]
                    \nonumber \\
		    &\!\!\!\!+&\!\!\!\!
	       \frac{f_{\pi}^2 e}{4}
               \bf{A}^{\rm T}\cdot\rm(\phi_{1}\nabla\phi_{2}-
                      \phi_{2}\nabla\phi_{1})
               +\frac{f_{\pi}^2 e^2}{8}(\bf A^{\rm T})^{\rm 2}\rm
                (\phi_{1}^{2}+\phi_{2}^{2}) \nonumber \\
	&\!\!\!\!-&\!\!\!\!\ 2 N e^2 \phi_3
	(\mbox{\boldmath $\pi$}^T + \frac{\nabla}{\nabla^2}\rho)
	\cdot \nabla\times(\phi_0 \bf A^{\rm T}\rm)+
	2 N^2 e^4 [\phi_3\nabla\times(\phi_0 
	\bf A^{\rm T}\rm)]^2 \nonumber \\ 
	 &\!\!\!\!-&\!\!\!\! 2 N^2 e^4 
	\frac{(\phi_0 \phi_3 \nabla \phi_0 \cdot \bf B\rm)^2}
	     {(\nabla \phi_0)^2}\;\;, \nonumber		
\end{eqnarray}
where $\phi_0 = (1-\phi_{i}^2)^{\frac{1}{2}} \; $.
We see that only the physical transverse components of the
vector potential enter in the expression of the Lagrangian density.
( Note that $ \bf B=\nabla\times A^{\rm T} $ ).
%
%%%%%%%%%%%%%%%%%%%%%%%%%%%%%%%%%%%%%%%%%%%%%%%%%%%%%%%%%%%%%%%%%%

%
%%%%%%%%%%%%%%%%%%%%%%%%%%%%%%%%%%%%%%%%%%%%%%%%%%%%%%%%%%%%%%%%%
%
\section{Conclusion}
\label{Con}

The two-flavour Wess-Zumino-Witten model coupled to electromagnetism 
is treated as a constrained system in the context of the 
Faddeev-Jackiw formalism. The treatment is complete since no expansion 
is used as in (10) where the U field is expanded in series of powers 
of the pion fields. The system has three true second class constraints. 
We write the Lagrangian density as an expression first order in time 
derivatives of the dynamical variables. Then by solving the constraints 
for some of the variables, substituting back into the Lagrangian 
density and performing Darboux's transformations we obtain a canonical 
expression with reduced coordinate but with the unphysical  
$\bf A^{\rm L}$ still occuring in the Hamiltonian density 
and specifically 
in the part which comes from the Wess-Zumino term. This is to be 
contrasted with the former case (10) where $\bf A^{\rm L}$ cancels 
out exactly when term up to second and third order in the Goldstone 
boson fields are kept. The existence of $\bf A^{\rm L}$ 
and $p_0$ in the 
Hamiltonian and not in the canonical one-form leads to a new 
set of constraint equations the solution of which fixes the value 
of $\bf A^{\rm L}$ and $p_0$. Finally we end up with unconstrained 
Coulomb-gauge canonical Lagrangian density (25) whose coordinate 
space consists of $\phi_i \;(i=1,2,3)\;,\; \bf A^{\rm T}$ and their 
conjugate momenta.  
%
%%%%%%%%%%%%%%%%%%%%%%%%%%%%%%%%%%%%%%%%%%%%%%%%%%%%%%%%%%%%%%%%%
%
\section{Appendix}
\label{app}

Our metric is $g_{\mu \nu}=diag(1,-1,-1,-1) \;\; , \;\;
Q=diag(2/3,-1/3)$
is the charge matrix,
 $D_\mu=\partial_\mu + ieA_\mu [Q,\;\;]$
denotes the covariant derivative.
By $\tau_{i}\hspace{0.25em},\hspace{0.25em} i=1,2,3$  we denote
Pauli matrices. We choose $ e>0 $ so that the electric charge of
the electron is $-e$. We define $ \epsilon^{0123}=1$ . 
By $\mbox{\boldmath $\pi$}$ we denote the electric field $\bf E $. 
Summation over repeated indices is understood.
%%%%%%%%%%%%%%%%%%%%%%%%%%%%%%%%%%%%%%%%%%%%%%%%%%%%%%%%%%%%%%%%%%%%

\end{document}